\documentstyle[epsf,epsfig,rotate]{mn}
\newcommand{\be}{\begin{equation}}
\newcommand{\ee}{\end{equation}}

\title
[ MHD turbulence]
 {The interaction of a giant planet with a disc with MHD turbulence I: 
The initial  turbulent disc models }
\author[J.C.B.Papaloizou \& R.P.Nelson]{John C.B. Papaloizou \&
Richard P. Nelson \\
Astronomy Unit, Queen Mary, University of London, Mile End Rd, London E1 4NS}

\date{Received/Accepted}

\begin{document}

\maketitle

\begin{abstract} 
This is the first of a series of papers aimed at developing and interpreting
simulations of protoplanets interacting with turbulent accretion discs.
In this first paper we study the turbulent disc models
prior to the introduction of a perturbing protoplanet.
We study cylindrical disc models
in which a central domain is in Keplerian rotation
and unstable to the magnetorotational instability (MRI).  
Models of varying disc size
and aspect ratio $H/r$ are considered with magnetic fields having
zero net flux. 
We relate the properties
of the turbulent models to classical viscous disc theory
(Shakura \& Sunyaev 1973). 
All models were found to attain a turbulent state in their
Keplerian domains  with
volume averaged stress parameter $\alpha \sim 5\times 10^{-3}.$
 At any particular time the vertically and azimuthally averaged
value  exhibited  large  fluctuations
in radius.
However, an additional  time average
 over periods exceeding $3$ orbital periods at
the outer boundary of the  Keplerian domain
 resulted in a more  smoothly varying quantity with radial
variations within a factor of two or so.

The vertically and azimuthally averaged radial velocity showed
much larger spatial and temporal fluctuations,
requiring  additional time averaging for  at least $7-8$ orbital periods
at the outer boundary of the Keplerian domain to
limit them.
Comparison with the value derived from the averaged stress
using viscous disc theory yielded schematic agreement
for feasible averaging times but with some indication
that the effects of residual fluctuations remained.

The behaviour described above must be borne in mind when
considering laminar disc simulations with anomalous Navier--Stokes viscosity.
This is because the operation of a viscosity
as in classical viscous disc theory
with anomalous viscosity coefficient 
cannot apply
to a turbulent disc undergoing rapid changes
due to external perturbation.  The classical theory
can only be used to describe the time averaged behaviour of the
parts of the disc that are in a statistically steady condition for long enough
for appropriate averaging to be carried out.

\end{abstract}


\section{Introduction}\label{S0} 
\noindent
The recent and ongoing discovery of extrasolar giant planets has stimulated
renewed interest in the theory of planet formation (e.g. Mayor \& Queloz 1995;
Marcy, Cochran, \& Mayor 1999; Vogt et al. 2002).
In particular the discovery of
giant planets close to their central stars has led to the idea
that they migrated inwards due to gravitational
interaction with the gaseous disc.
Previous studies of the interaction between a protoplanet and a  laminar
but viscous
disc [Papaloizou \& Lin (1984); Lin \& Papaloizou (1986, 1993);
Bryden et al. (1999);
Nelson et al. (2000); D'Angelo, Henning \& Kley (2002)]
indicate  that a
protoplanet in the Jovian mass range will open a gap and that the torques
exerted through the disc protoplanet interaction can produce inward orbital
migration. 
However, the effect of the turbulence producing the anomalous viscosity
has yet to be taken into account. The origin of this turbulence
was uncertain until Balbus \& Hawley (1991) provided an explanation
for its origin through the operation of the magnetorotational instability 
(MRI).
Improved computational resources now makes it feasible to consider
three dimensional simulations of turbulent discs interacting with protoplanets.

This is the first of a series of papers aimed at developing and interpreting
such simulations. In this first paper we focus on the turbulent disc models
prior to the introduction of a perturbing protoplanet.
The effect of introducing the protoplanet will
be considered in a companion paper (Nelson \& Papaloizou (2002) -- hereafter 
paper II).
To ease computational requirements we adopt cylindrical disc models
with no vertical stratification. For this first study we assume that the disc
is adequately ionized throughout so that ideal MHD
applies and  consider models with zero net magnetic flux.
 We extend an initial study by Steinacker \& Papaloizou (2002)
(hereafter SP) to consider a wider range of models with varying
disc size aspect ratio and resolution in the context of classical
`$\alpha$' viscous disc theory (Shakura \& Sunyaev 1973).
This is because in addition to providing the most common and simple conceptual
framework for disc modeling it still 
remains the main contact with observations
(Balbus \& Papaloizou 1999).
We here focus on time averages of disc quantities such as the
$r \phi$ stress and radial inflow velocity and study the extent to which
stable behaviour of these quantities relates to classical disc
theory.




The plan of the paper is as follows.
We describe the basic equations and model set up
in \S \ref{S1}. We describe  the  numerical procedure  
in \S \ref{S2}  and   our  numerical results in \S  \ref{S3a}.
Finally, we summarize our results in \S \ref{S6}.

\section{Initial model setup} \label{S1} 
\noindent
The governing equations for MHD written in a frame rotating
with uniform  angular velocity $\Omega_p {\bf {\hat k}} $
with $ {\bf {\hat k}} $ being the unit vector in the vertical direction
are:
\begin{equation}
\frac{\partial \rho}{\partial t}+ \nabla \cdot {\rho\bf v}=0, \label{cont}
\end{equation}
\begin{eqnarray}
\rho \left(\frac{\partial {\bf v}}{\partial t}
 + {\bf v}\cdot\nabla{\bf v}\right) + 2\Omega_p 
{\bf {\hat k}}{\bf \times}{\bf v} & = &
 - \nabla p -\rho \nabla\Phi \nonumber \\
&&
+ \frac{1}{4\pi}(\nabla \times {\bf B}) \times {\bf B}, \label{mot}
\end{eqnarray}
\begin{equation}
\frac{\partial {\bf B}}{\partial t}=\nabla \times ({\bf v} \times {\bf B}).
\label{induct}
\end{equation}
where ${\bf v}, P, \rho, {\bf B}$ and $\Phi$ denote the fluid
velocity, pressure, density, magnetic field, and potential, respectively.
$\Phi$ contains contributions due to gravity
and the centrifugal potential $-(1/2)\Omega_p^2 r^2.$
\noindent
We use a  locally isothermal equation of state
\begin{equation}
P(r)= c(r)^2 \cdot \rho,
\end{equation}
where $c(r)$ denotes the sound speed
which is specified as a fixed function of $r.$

\noindent
The  models investigated may be described
as cylindrical discs (e.g. Hawley 2001).
Adopting cylindrical coordinates $(z,r,\phi),$ the gravitational
potential is taken to depend on  $r$ alone, so that
$\Phi=-GM/r - (1/2)\Omega_p^2 r^2 $, where
$M$ is the central mass, $G$ is the gravitational constant, and the second term
represents the centrifugal potential where applicable.
Thus the cylindrical disc models do not include a full
treatment of the disc vertical structure. Models of this type are employed
due to the high computational overhead that would be required
to resolve fully the disc vertical structure of a stratified model. 

In this paper we consider the time dependent evolution
and turbulent state that is set up in five models that we label as 
A, B, C, D, and E.
These are all initiated with zero net magnetic flux in both 
vertical and azimuthal directions which is conserved for the duration
of the simulations. As in SP periodic boundary conditions were used in the
vertical and azimuthal directions, and each of the models has a radial
computational domain $(r_{1}, r_{2})$ in which is embedded a central
Keplerian domain $(r_{a1}, r_{a2})$ where the
angular velocity $\Omega \propto r^{-3/2}$ and 
which becomes unstable due to the MRI. Interior and exterior
to this Keplerian domain, adjacent to the inner and outer rigid
radial boundaries,
there exist regions in which the angular
velocity profile is non Keplerian. These are described below.
The relevant model parameters are given in table \ref{table1}.
Dimensionless units of length and time
are adopted such that $r_{1} =1 $ and $GM=1.$

The angular velocity profile is chosen to be stable to the MRI
in the boundary domains (where $r_1 \le r < r_{a1}$ and $r_{a2} < r \le r_2$),
the inner one of which can be thought of 
as modeling the boundary layer between star and disc.
For the models considered here, we adopted  $c^2(r) \propto 1/r$
with constant of proportionality such that 
$c/(r\Omega) = 0.1$ (models A, B, E) or $c/(r\Omega) = 0.2$ (models C, D),
apart from in the inner boundary domain where $c^2(r) \propto r^{-5/4}.$
In the Keplerian domain the initial density was such that
$\rho \propto 1/r$ for all models. In models A, C, and D  the
angular velocity was initially constant in both boundary domains.
In models B and E we took $\Omega \propto r^2$ in the inner boundary domain.
Models A, C, and D were initiated with initial toroidal fields 
contained within the Keplerian
domain while models B and E were initiated with poloidal fields.
As in  SP
the initial  magnetic field  was given by
\begin{equation}
{\bf B}_i=B_0\sin\left(2 n_R\pi \frac{r-r_m}{r_{m1}-r_m}\right)\bf{e}_i, 
\label{Binit}
\end{equation}
with the integer $n_r$ being the number of $2\pi$ cycles.  
The index $i$ indicates either the vertical or the toroidal field component
with the corresponding unit vector $\bf{e}_i.$  For toroidal fields
$B_0$ is  constant while for
vertical fields $B_0 \propto 1/r.$
The magnetic field was applied in an annulus within the Keplerian
domain with  inner and outer  
bounding radii  $r_m$ and $r_{m1}$ respectively.
The normalization of $B_0$ was chosen such that the initial
magnetic energy in the
Keplerian domain
expressed in units of the volume integrated pressure there
was $0.03$ for models A, C, and D, $0.003$ for
model B, and
$0.002$ for model E.
As the calculations proceed the
magnetic field is seen to diffuse throughout the Keplerian domain until it is
more or less 
fully magnetised, leading to a final turbulent state that is described in 
subsequent sections.
 
Model E was initiated in the inertial frame with an azimuthal domain of extent 
$\pi/3$, and is the disc model used to study disc-planet interaction in 
paper II.
This was run up to time $t=3825.5$ when it had attained a 
fully turbulent state.
For the purpose of studying the interaction between a turbulent protostellar
disc and an embedded protoplanet, the azimuthal domain for this 
model was then extended to
$2\pi$ by stacking
six of the $\pi/3$ sectors together. A transformation to a rotating frame with
$ \Omega_p = 0.30645$ was carried out and the evolution continued.
Some results obtained at this stage are described below.
We found no significant dependence of the evolution of the models
on $\Omega_p$ or the extent of the $\phi$ domain once the latter
exceeded $\pi/3.$

 \begin{table*}
 \begin{center}
 \begin{tabular}{|l|l|l|l|l|l|l|l|l|l|l|l|l|l|}\hline\hline
       &     &     &     &     &        &        &        &     &     &   &    &    & \\
 Model&$z_1$&$z_2$&$r_1$&$r_2$&$\phi_2$&$r_{a1}$&$r_{a2}$&$n_z$&$n_r$&$n_{\phi}$ &$r_m$&$r_{m1}$&$n_r$\\
       &     &     &     &     &        &        &        &     &     &   &    &    &  \\
 \hline
 \hline
 A&$-0.2$& $0.2$ & $1.0$ & $6.1$&  $\pi/3$&$1.25 $&$5.0  $&$40$&$380$&$100$&$2.35$&$4.35$&$ 2$\\
 B&$-0.2$& $0.2$ & $1.0$ & $4.0$&  $\pi/2$&$1.2  $&$3.7  $&$54$&$334$&$108$ &$1.33$&$3.33$&$ 6$\\
 C&$-0.3$& $0.3$ & $1.0$ & $4.5$&  $\pi/3$&$1.5  $&$3.7  $&$44$&$334$&$100$&$2.33$&$3.33$&$ 1$\\
 D&$-0.3$& $0.3$ & $1.0$ & $4.5$&  $\pi/3$&$1.5  $&$3.7  $&$44$&$334$&$150$&$2.33$&$3.33$&$ 1$\\
 E&$-0.3$& $0.3$ & $1.0$ & $8.8$&  $\pi/3$&$1.2  $&$7.2  $&$60$&$370$&$100$&$3.5$&$6.5$&$ 3$ \\
\hline
\end{tabular}
\end{center}
\caption{ \label{table1}
The first column gives the model label, the second and third  
give the  vertical domain of extent $L_z,$
the fourth and fifth give the radial domain while the sixth column
gives the maximum extent of the azimuthal domain.
The Keplerian domain is specified in columns seven and eight
and the numbers of equally spaced grid points in the vertical, radial
and azimuthal directions  are given in  columns nine, ten and eleven.
The final three columns give the boundaries of the domain in which
the initial magnetic field was applied and the number of $2\pi$ cycles in the
functional form.}
\end{table*}
\noindent

\subsection{Numerical procedure} \label{S2}
The numerical scheme that we employ is 
based on a spatially second--order accurate method that computes the
advection using the monotonic transport algorithm (Van Leer 1977).
The MHD section of the code uses the
method of characteristics
constrained transport (MOCCT) as outlined in Hawley \& Stone (1995)
and implemented in the ZEUS code.
The code has been developed from a version
of NIRVANA originally written by U. Ziegler
(see SP, Ziegler \& R\"udiger (2000), and
references therein)

\section{Numerical Results} \label{S3a}
We now present numerical results for the evolution
of models A -- E into a saturated turbulent state
and characterize their average evolutionary behaviour.
To do this we use spatial and temporal averages
of the state variables as indicated below.
 
\subsection{Vertically and horizontally averaged stresses
 and angular momentum transport}\label{S3}
\noindent
In order to describe average properties of the turbulent models, we 
use quantities that are both
vertically and azimuthally averaged over the $(\phi, z)$ domain 
(e.g. Hawley 2000)
and in some cases an additional time average.
The vertical and horizontal average of
$Q$ is defined through
\begin{equation}
{\overline {Q(r,t)}} ={\int \rho  Q dz d\phi \over \int  \rho dz d\phi}.
\end{equation}                                

Below the average is taken over the full $2\pi$ in azimuth.
If a smaller domain is used  $2\pi$ should be replaced by the extent
of this domain in what follows below in this section. In practice we have
found that the results are independent of the size of the $\phi$ domain,
with the smallest azimuthal domain that we have considered being
$\pi/3.$

\noindent The vertically and azimuthally averaged
continuity equation (\ref{cont}) may  then be written
\begin{equation}
\frac{\partial \Sigma}{\partial t}+\frac{1}{r}
\frac{\partial\left(r\Sigma\overline{v_r}\right)}{\partial r}=0, \label{cona}
\end{equation}
where
\noindent the  disc surface density is given by
\begin{equation}
\Sigma = {1\over 2\pi}\int \rho dz d\phi.
\end{equation}
(Note that in performing the averaging the basic equations
are integrated over the vertical and azimuthal domains without
first multiplying by the density.)

\noindent  The  vertically and azimuthally
averaged Maxwell and
Reynolds stresses,  are  respectively defined as follows:
\begin{equation}
T_M(r,t)=2\pi
\Sigma{\overline{\left({B_r(z,r,\phi,t) B_\phi(z,r,\phi,t) \over 4\pi\rho}\right)}}
\end{equation}
and
\begin{equation}
T_{Re}(r,t)=2\pi
\Sigma
{\overline{\delta v_r(z,r,\phi,t)\delta v_\phi(z,r,\phi,t)}}.
\end{equation}
 Here the velocity fluctuations $\delta v_r$ and $\delta v_\phi$
are defined through,
\begin{equation}
\delta v_r(z,r,\phi,t)=v_r(z,r,\phi,t)-{\overline{v_r}}(r,t),
\end{equation}
\begin{equation}
\delta v_\phi(z,r,\phi,t)=v_\phi(z,r,\phi,t)- {\overline{v_{\phi}}}(r,t).
\end{equation}
The Shakura  \& Sunyaev (1973) 
$\alpha$ stress parameter appropriate to the  
total stress is  defined by 
\begin{equation}
\alpha(r,t)=\frac{T_{Re}-T_M}{2\pi
\Sigma{\overline{ \left(P/\rho\right)}}},
\label{alphaeqn}
\end{equation}
The vertically and azimuthally averaged azimuthal
component of the equation of motion (\ref{mot}) 
can be written in the form
\begin{equation}
\frac{\partial \left(\Sigma \overline{j}\right)}{\partial t}
+\frac{1}{r}\left(
\frac{\partial\left( r\Sigma\overline{v_r}\overline{j}\right)}{\partial r}
+\frac{\partial\left(\Sigma r^2\alpha\overline{P /\rho}\right)}{\partial r}
\right) =0.\label{mota}
\end{equation}                 
[see Balbus \& Papaloizou (1999)].
Here $j=rv_{\phi}$ is the specific angular momentum.

\noindent Using equation (\ref{cona}), equation(\ref{mota})  may also be written as
\begin{equation}
\Sigma r \left(\frac{\partial  \overline{j}}{\partial t} +
\overline{v_r}\frac{\partial \overline{j}}{\partial r}\right)
=-\frac{\partial\left(\Sigma r^2\alpha\overline{P /\rho}\right)}{\partial r}.
\label{motab}
\end{equation}
Note too that the averages may be extended without change
of formalism  to incorporate 
a time average such that for any quantity
\begin{equation}
{\overline {Q(r,t)}} \rightarrow  
{1\over 2\Delta} \int^{t+\Delta}_{t-\Delta}{\overline {Q(r,t')}}dt',
\label{motac}
\end{equation}
where the time average is carried out over an interval $2\Delta$
centered on the  time $t$ and this is incorporated
into the definition of $\Sigma.$

Under quasi-steady state conditions in the mean,
if the time average is carried out over a 
sufficiently long interval,  
we should be able to neglect the time variation of
the mean specific angular momentum ${\overline j}.$
Then, as in the normal viscous disc theory, 
we expect to  have an expression for the mean radial velocity
of the form
\begin{equation}
\Sigma r 
\overline{v_r} \frac{\partial \overline{j}}{\partial r}
=-\frac{\partial\left(\Sigma r^2\alpha\overline{P /\rho}\right)}{\partial r}.
\label{motad}
\end{equation}

\subsection{Model A}
\begin{figure}
\centerline{
\epsfig{file= 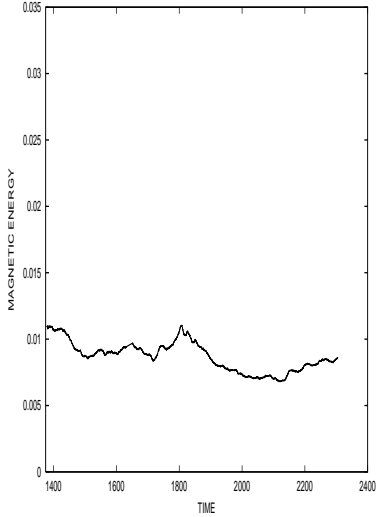,height=7.cm,width=5.cm,angle =270} }
\caption[]{ Magnetic energy in the Keplerian domain
expressed in units of the volume integrated pressure
as a function of time
for model A for $t > 1378.6.$
This model was run for up to $2300$ time units.
This corresponds to $366$ orbits at $r=1,$
$70$ orbits at $r=3,$ and
$33$ orbits at $r=5.$}
\label{fig1}
\end{figure}

The time dependent evolution of the total  magnetic energy in the 
Keplerian domain
expressed in units of the volume integrated pressure,
$ \int_V({\bf B}^2/8\pi) dV /  \int_V P dV, $
is plotted as a function of time
for model A in figure \ref{fig1}.
This model was run for up to $2300$ time units.
This corresponds to $366$ orbits at $r=1,$
$70$ orbits at $r=3,$ and
$33$ orbits at $r=5.$ The initial value of
the ratio of the total  magnetic energy to volume integrated pressure
$1/\langle \beta \rangle =\int {\bf B}^2/(8\pi)dV / \int PdV$
was $0.03$ for model  A and all others  initiated with toroidal fields.
After the onset of the MRI, some loss of the rather high initial magnetic
energy due to reconnection occurs, and a relaxed turbulent state is attained
after about ten orbits at the outer boundary of the Keplerian
domain. The statistical properties of this do not depend
on the initial conditions for models with zero net flux
(see SP and below). 

\noindent
\begin{figure}
\centerline{
\epsfig{file=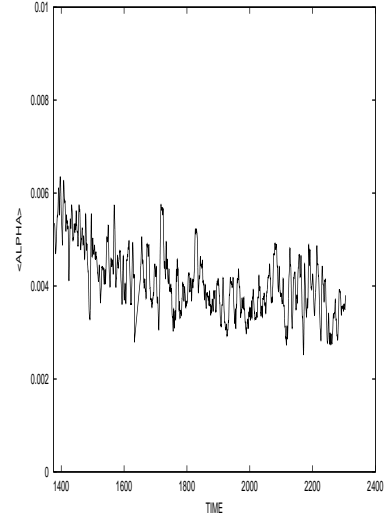,height=7.cm,width=5.cm,angle =270} }
\caption[]{The stress parameter
$\alpha$ 
volume averaged over the Keplerian domain is plotted
as a function of time for model A.}
\label{fig2}
\end{figure}
\noindent
The stress parameter
$\alpha$, defined in equation (\ref{alphaeqn}),
volume averaged over the Keplerian domain is plotted
as a function of time for model A in figure \ref{fig2}.
Mean values are typically $\sim 0.004.$

\begin{figure}
\centerline{
\epsfig{file=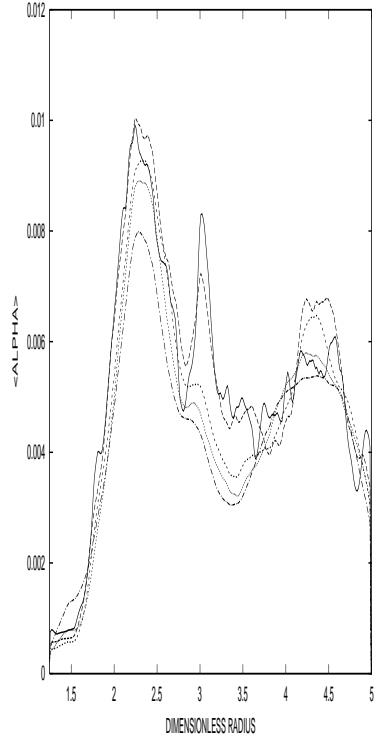,height=10.cm,width=5.cm,angle =270} }
\caption[]
{ Time averages of the  stress parameter
 $\alpha$ are plotted as a function of dimensionless radius
for model A. These end at times $t=1391.6$ solid line,
$t=1430.0$ dashed line, $t=1549.7$ short dashed line
$t=1646.4$ dotted line and $t= 1841.6$ dot  dashed line respectively.
In each case the time averaging starts at $t=1378.6.$
}
\label{fig3}
\end{figure}

We now examine how the stress parameter $\alpha$ varies in space and time.
To do this we consider time averages.
Time averages of the  stress parameter
$\alpha$ are plotted as a function of dimensionless radius
for model A in figure \ref{fig3}. Time averaging ends at times $t=1391.6$,
$t=1430.0$, $t=1549.7$,
$t=1646.4$ and $t= 1841.6$, respectively.
In each case the time averaging starts at $t=1378.6.$
Although a snapshot of $\alpha$ may reveal quite large variations
(see figure \ref{fig13} below for model C, and also SP)
after quite a short time a stable picture emerges.
This is apparent after time averaging over an interval
as short as $12$ units, which represents 2 orbits at the inner boundary
and only $0.2$ orbits at the outer boundary. 
Although from figure~\ref{fig3} there is still some erratic behaviour
visible for an averaging of $50$ time units, 
for averaging periods exceeding $70$ units, or one orbital
period at the outer boundary of the active domain,
the time averaged stress parameter appears
to be a reasonably smooth function of $r.$
The variation of the time averaged $\alpha$ in the active
domain is between $0.008$ and $0.003.$
These values are typical of those seen in local
shearing box simulations starting
with magnetic fields with zero net flux
[Hawley, Gammie \& Balbus (1996); Brandenburg et al. (1996); 
Brandenburg (1998);
Fleming, Stone \& Hawley (2000)].
Smaller values are obtained near the inner boundary.
This is probably because magnetic field
has not yet diffused to the inner boundary 
so that a long term turbulent steady state has not been reached there
(note that the initial magnetic field was applied in an annulus away
from the boundary domains as described in section~\ref{S1}).

In general the time averaged stress parameter $\alpha,$ and
because of the relatively small fluctuations in time 
of $\Sigma$, also 
the time averaged stress itself attains a stable pattern
after only a short period of time averaging.

On the other hand, a stable pattern for the time averaged radial velocity
takes significantly longer to attain. This is because from viscous disc
theory we expect a characteristic value 
$\sim 1.5\alpha (H/r)^2  (r\Omega)\sim 8\times 10^{-5}r^{-1/2}.$
A snapshot of the $v_r$ averaged over the vertical and azimuthal domain is
typically
between one and two order of magnitudes higher, corresponding to large temporal 
fluctuations in that
quantity (see also SP).

\begin{figure}
\centerline{
\epsfig{file=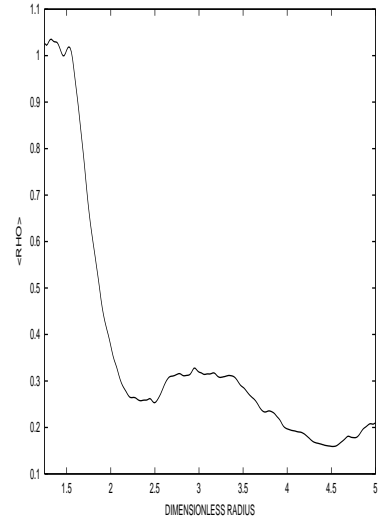,height=7.cm,width=5.cm,angle =270} }
\caption[]
{ A snapshot of $\Sigma /L_z$ is  
 plotted as a function of radius for model A 
at the specific  time $t=1841.6$}
\label{fig3a}
\end{figure}

\begin{figure}
\centerline{
\epsfig{file= 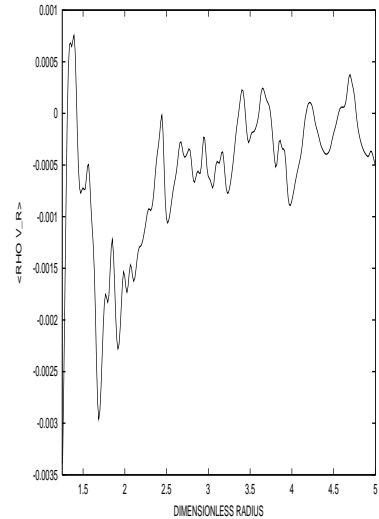,height=7.cm,width=5.cm,angle =270} }
\caption[]{
A snapshot  of the product of the  vertically and azimuthally averaged
values of $v_r$ with  $\Sigma /L_z,$
 is plotted as a function of dimensionless radius
for model A
at time $t=1841.6.$
}
\label{fig4}
\end{figure}

We present a snapshot of $\Sigma /L_z = 1/(2\pi L_z) \int \rho d\phi dz$ 
plotted as a function of radius for model A
at the specific  time $t=1841.6$ in figure \ref{fig3a} (where $L_z$ is the
vertical extent of the computational domain).
The temporal fluctuations in this quantity are found to be small,
and representing the density averaged
over the azimuthal and vertical domains it indicates some mass accretion
towards the inner regions (see also figure~\ref{fig21} for model E below).
A snapshot of the product of the vertically and azimuthally averaged
values of $v_r$ with  $\Sigma /L_z,$
is plotted as a function of dimensionless radius
for model A
at time $t=1841.6$ in figure \ref{fig4}. This quantity is related to the
instantaneous radial mass flux.
Characteristic values of the averaged $v_r$
are seen to be much larger than those  expected 
from classical viscous disc theory
which are recovered only after a long period of time averaging.
 
\begin{figure}
\centerline{
\epsfig{file=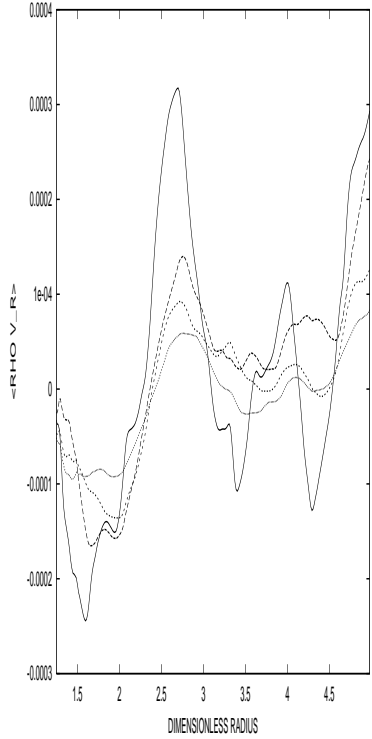,height=10.cm,width=5.cm,angle =270} }
\caption[]
{ The product of the time averages of the vertically and azimuthally averaged
values of $v_r$ with the time averaged $\Sigma /L_z,$
 are plotted as a function of dimensionless radius
for model A. These are taken at times $t=1430.0$ (solid line),
$t=1549.7$ (dashed line), $t=1646.4$ (short dashed line) and
$t=1841.6$ (dotted line).
The time averaging starts from $t=1378.6.$
}
\label{fig5}
\end{figure}

To show this, we plot the product of the time averages 
of the vertically and azimuthally averaged
values of $v_r$ with the time averaged $\Sigma /L_z,$
as a function of dimensionless radius in figure \ref{fig5}
for model A. The  averages end at times $t=1430.0$,
$t=1549.7$, $t=1646.4$  and
$t=1841.6$.
The time averaging starts at $t=1378.6.$
Even the longest two averages over $463$ and $268$ time units,
although indicating
magnitudes comparable to those expected from viscous disc theory, 
deviate significantly. The shortest average over $52$ time units
is very different in character. From this we conclude that periods
of time up to $7$ orbital periods at the outer boundary
of the active domain are required to obtain radial velocities that
can be compared with viscous disc theory, and 
then the comparison can be  moderately successful
for the times over which averages can be taken here.

\begin{figure}
\centerline{
\epsfig{file= 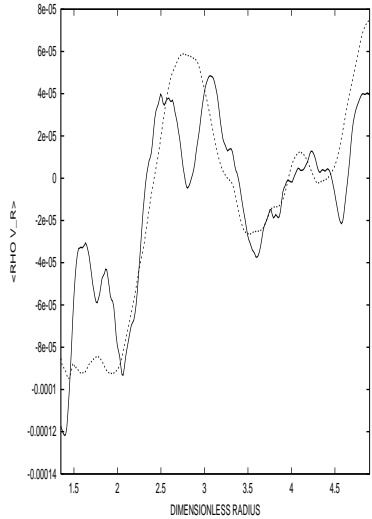,height=7.cm,width=5.cm,angle =270} }
\caption[]
{ The product of the time averages of the vertically and azimuthally averaged
values of $v_r$ with the time averaged $\Sigma /L_z,$
 is plotted as a function of dimensionless radius
for model A in (dotted line). The time average 
starts at $t=1378.6$ and ends at
$t= 1841.6$.
The value obtained from the time averaged stress using
equation(\ref{motad}) is also plotted (solid line). The method used for doing
this is described in the text.
}
\label{fig6}
\end{figure}

We illustrate this by plotting
the longest time average of the vertically and azimuthally averaged
values of $v_r$ with the time averaged $\Sigma /L_z,$
as a function of dimensionless radius
for model A in figure \ref{fig6}.
The value of this quantity that we obtained from   
equation (\ref{motad}), using the time averaged stress
$\alpha {\bar P/\rho}$ obtained in the simulation, 
the Keplerian value for ${\overline j}$,
and using numerical differentiation is also plotted as the solid line.
This latter quantity varies somewhat erratically
because of the numerical differentiation, but
nonetheless the general agreement is reasonable.

\subsection {Model B}

\begin{figure}
\centerline{
\epsfig{file= 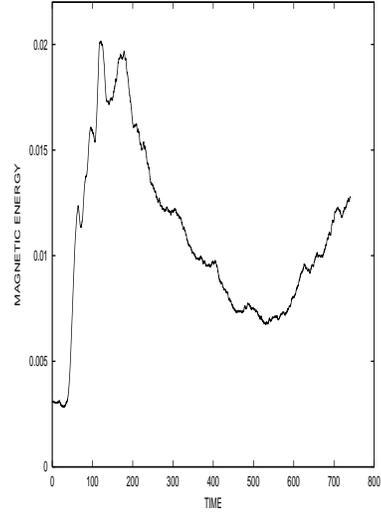,height=7.cm,width=5.cm,angle =270}}
\caption[]
{Magnetic energy in the Keplerian domain
expressed in units of the volume integrated pressure
as a function of time
for model B.
}
\label{fig7}
\end{figure}

It is expected that models with 
initially zero net flux should
attain similar steady states independent of initial conditions,
disc thickness and size. To investigate this we 
considered model B which started from a vertical field
and had outer boundary of the active domain at $r=3.7.$
The magnetic energy in the Keplerian domain
expressed in units of the volume integrated pressure
is plotted as a function of time
for model B in figure \ref{fig7}.
As for model A and all others to be presented, this settles down 
to a turbulent state with $1/\langle \beta \rangle \sim 0.01.$

 \begin{figure}
 \centerline{
 \epsfig{file=  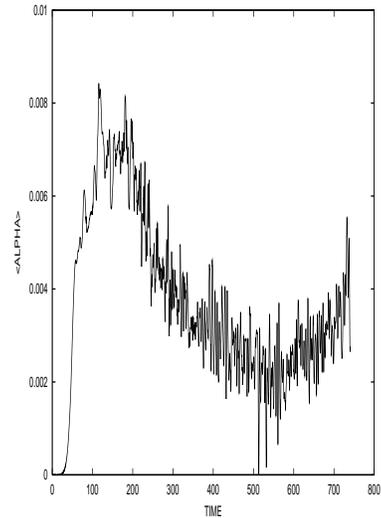 ,height=7.cm,width=5.cm,angle =270}}
 \caption[]
 {  The   stress parameter
 $\alpha$ volume averaged over the Keplerian domain is plotted
as a function of time for model B.
 }
 \label{fig8}
 \end{figure}

The stress parameter
$\alpha$, volume averaged over the Keplerian domain, is plotted
as a function of time for model B in figure \ref{fig8}.
This also takes on similar values and has similar behaviour  to other models.

\begin{figure}
\centerline{
\epsfig{file= 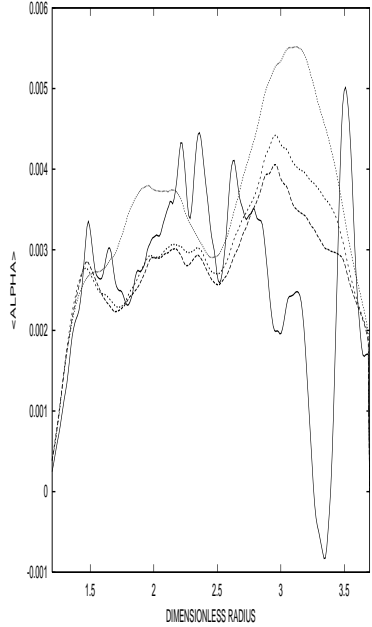 ,height=8.5cm,width=5.cm,angle =270} }
\caption[]{ Time averages of the 
 stress parameter $\alpha$ are plotted as a function of dimensionless radius
for model B. These  end at times $t=525.4$ solid line,
$t=634.5$ dashed line, $t=664.3$ short dashed line
and $t=740.6$ dotted line.
The time averaging starts at $t=515.2.$
}
\label{fig9}
\end{figure}

The behaviour of the time averages of the
stress parameter $\alpha$ are plotted as a function of dimensionless radius
for model B in figure~\ref{fig9}. These end at times $t=525.4$,
$t=634.5$, $t=664.3$ 
and $t=740.6$.
The time averaging starts at $t=515.2.$
The shortest average over only  $10$ time units, gives erratic behaviour 
and even  negative values.
However, for those taken over more than $120$ units a smooth stable
behaviour is obtained with typical values of $0.004$.  
But note that there are also long term cyclic trends
which are seen in the time averages (see also Brandenburg 1998). 

\subsection{ Models C and D}
We now describe models C and D.
These models had an increased sound speed such that $c/(r\Omega) =0.2.$
Further, model D had $50$ percent larger resolution in azimuth 
than model C in order to test the effects of azimuthal resolution. 

\begin{figure}
\centerline{
\epsfig{file= 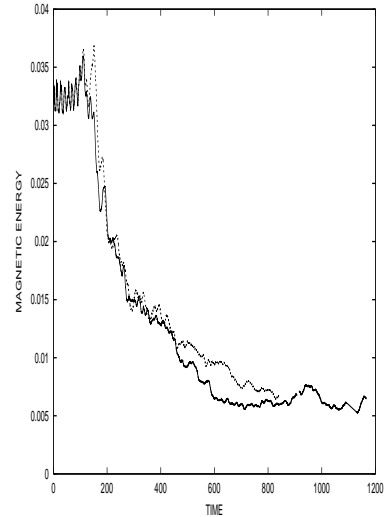,height=7.cm,width=5.cm,angle =270} }
\caption[]{
Magnetic energy in the Keplerian domain
expressed in units of the volume integrated pressure
as a function of time
for model C solid line and model D dashed line.
}
\label{fig10}
\end{figure}

The magnetic energy in the Keplerian domain
expressed in units of the volume integrated pressure
is shown as a function of time
for models C  and  D  in figure \ref{fig10}.
The behaviour of these models is very similar
and again we find  typical values $\sim 0.01$ for this quantity.
\noindent

\begin{figure}
\centerline{
\epsfig{file=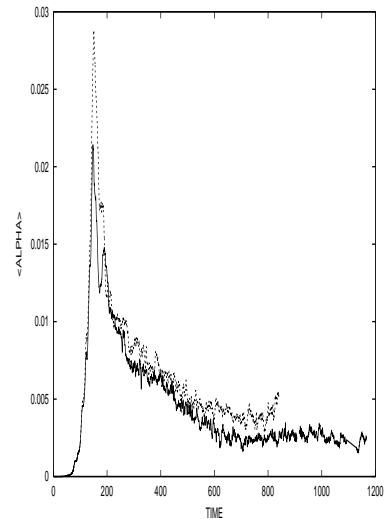,height=7.cm,width=5.cm,angle =270} }
\caption[]
{
Volume averages of the stress parameter $\alpha$ in the Keplerian domain
as a function of time
for model C solid line and model D dashed line.
}
\label{fig11}
\end{figure}

Volume averages of the stress parameter $\alpha$ in the Keplerian domain
for  these models are plotted as a function of time in figure \ref{fig11}.
Again these models behave similarly.
\begin{figure}
\centerline{
\epsfig{file= 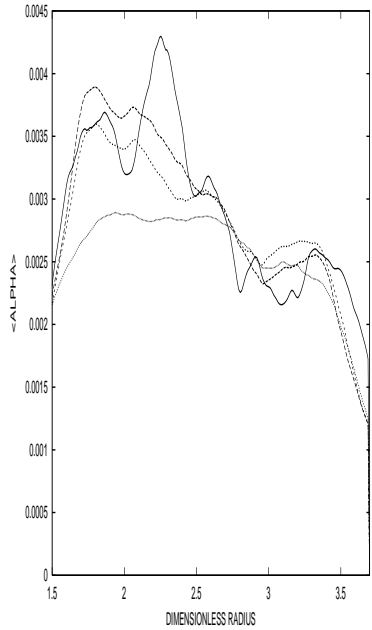,height=8.5cm,width=5.cm,angle =270} }
\caption[]
{Time averages of the  stress parameter
 $\alpha$ are plotted as a function of dimensionless radius
for model C. These are taken at times $t=932.6$ solid line,
$t=983.0$ dashed line, $t=1024.8$ short dashed line
and $t=1163.4$ dotted line.
The time averaging starts from $t=915.9.$
}
\label{fig12}
\end{figure}
\begin{figure}
\centerline{
\epsfig{file= 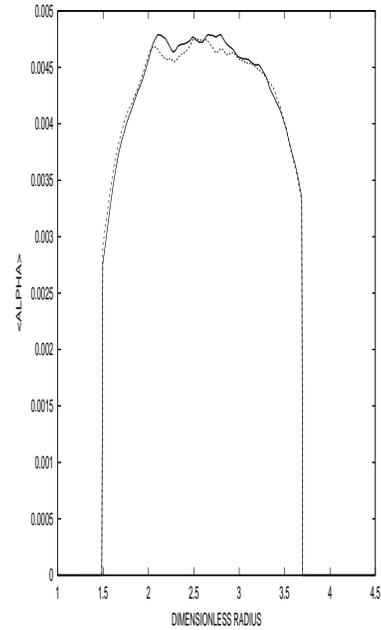,height=8.5cm,width=5.cm,angle =270} }
\caption[]
{Time averages of the  stress parameter
 $\alpha$ are plotted as a function of dimensionless radius
for model D. These are taken at times $t= 833.6$ solid line,
and $t=815.3$ dashed line.
The time averaging starts from $t= 447.5.$}
\label{fig12a}
\end{figure}

We plot time averages of the  stress parameter
 $\alpha$ as a function of dimensionless radius
for model C in figure \ref{fig12}. These are taken at times $t=932.6$,
$t=983.0$, $t=1024.8$
and $t=1163.4$.
The time averaging starts from $t=915.9.$
 A similar plot for model  D is given in figure \ref{fig12a}.
Here the averages are taken at times $t= 833.6$ solid line,
and $t=815.3$ dashed line.
The time averaging starts from $t= 447.5.$

The  time averages of 
 $\alpha$  are more uniform in  the thicker disc models C and D
with stable behaviour being attained for averaging periods exceeding
$4$ orbital periods at the outer boundary of the Keplerian domain.
The more uniform behaviour may occur because  the shorter viscous time
in these cases enables a better relaxation of the disc to a state
where memory of initial conditions is lost.

\begin{figure}
\centerline{
\epsfig{file= 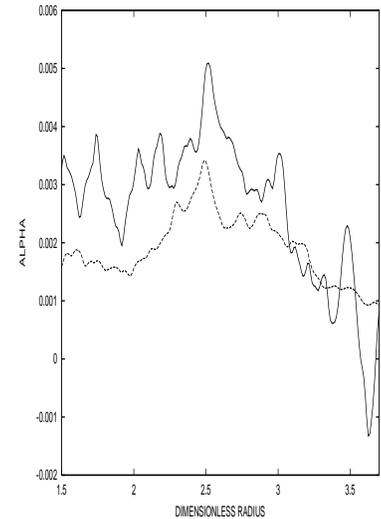 ,height=7.cm,width=5.cm,angle =270}}
\caption[]
{A snapshot of the  stress parameter $\alpha$
plotted as a function of radius for model C
at time $t=1163.4$
The solid curve corresponds to the total stress
while the lower dashed curve is obtained when only the magnetic
stresses are taken into account.
}
\label{fig13}
\end{figure}

To compare with the time averages a 
snapshot of the stress parameter $\alpha$
plotted as a function of radius for model C
at time $t=1163.4$ is shown in figure \ref{fig13}.

\begin{figure}
\centerline{
\epsfig{file=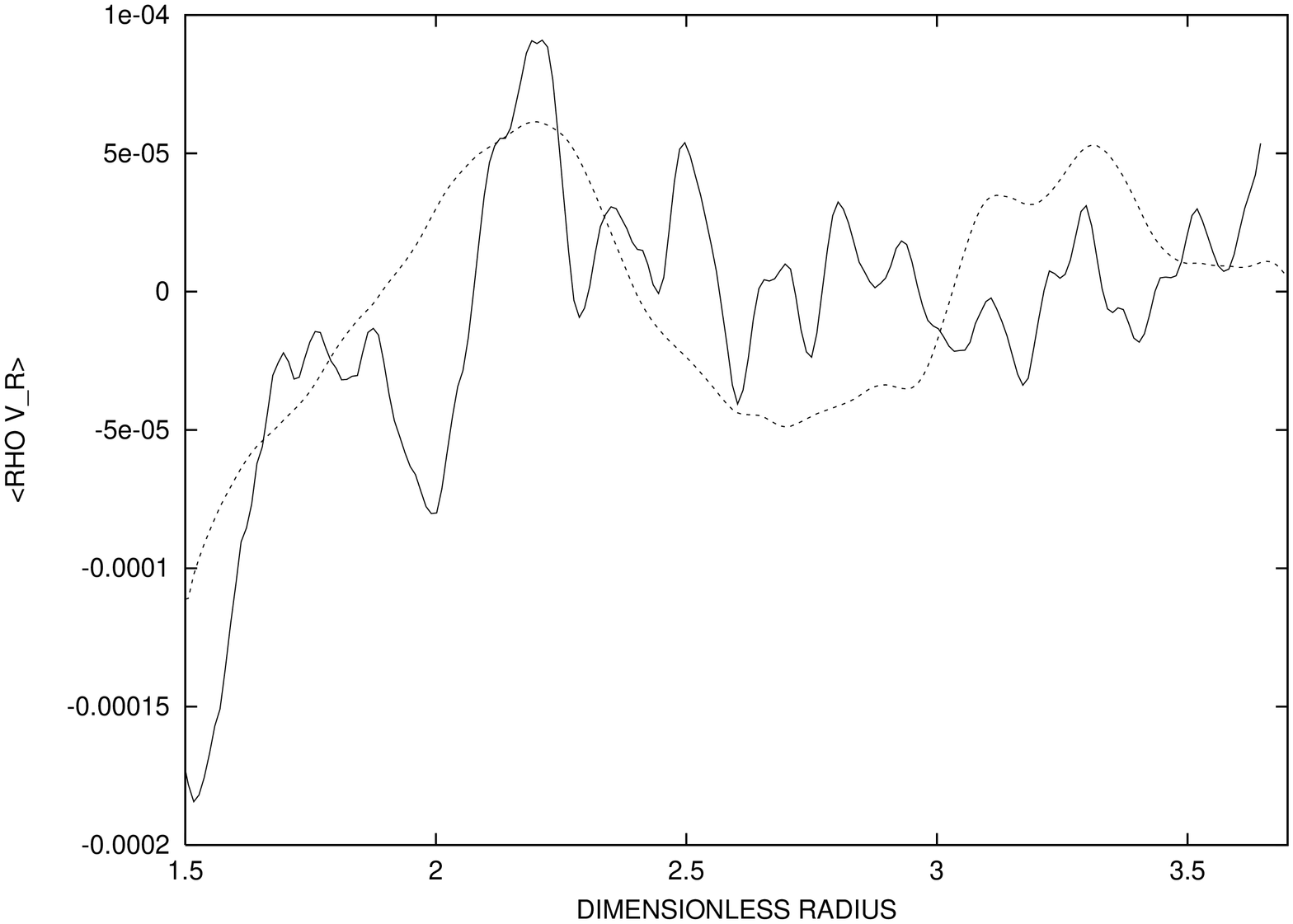,height=10.cm,width=5.cm,angle =270} }
\caption[]
{ The product of the time averages of the vertically and azimuthally averaged
values of $v_r$ with the time averaged $\Sigma /L_z,$
 is plotted as a function of dimensionless radius
for model D  (dashed line). The time average
starts at $t=447.5.$ and ends at
$t= 833.6$
The value obtained from the  time averaged stress using
 equation(\ref{motad}) is also plotted (solid line).
}
\label{fig51}
\end{figure}

To investigate the applicability of equation (\ref{motad}), we  plot
the  time average of the vertically and azimuthally averaged
values of $v_r$ with the time averaged $\Sigma /L_z,$
as a function of dimensionless radius
for model D in figure \ref{fig51}.
The value of this quantity that we obtained from
equation (\ref{motad}), using the time averaged stress,
 the Keplerian value for ${\overline j}$
 and using numerical differentiation is also plotted as the solid line.
As found for model A this quantity varies somewhat erratically
because of the numerical differentiation, but
there is schematic agreement.
 The time average
 was taken over eight orbital periods at the outer boundary
of the Keplerian domain which is not enough to reduce
the fluctuations in the radial velocity which can be up to
two orders of magnitude larger than the mean to a very low level.
This may require much longer averaging times comparable to the viscous timescale
which are not feasible to consider here.

\subsection{Model E}

\begin{figure}
\centerline{
\epsfig{file=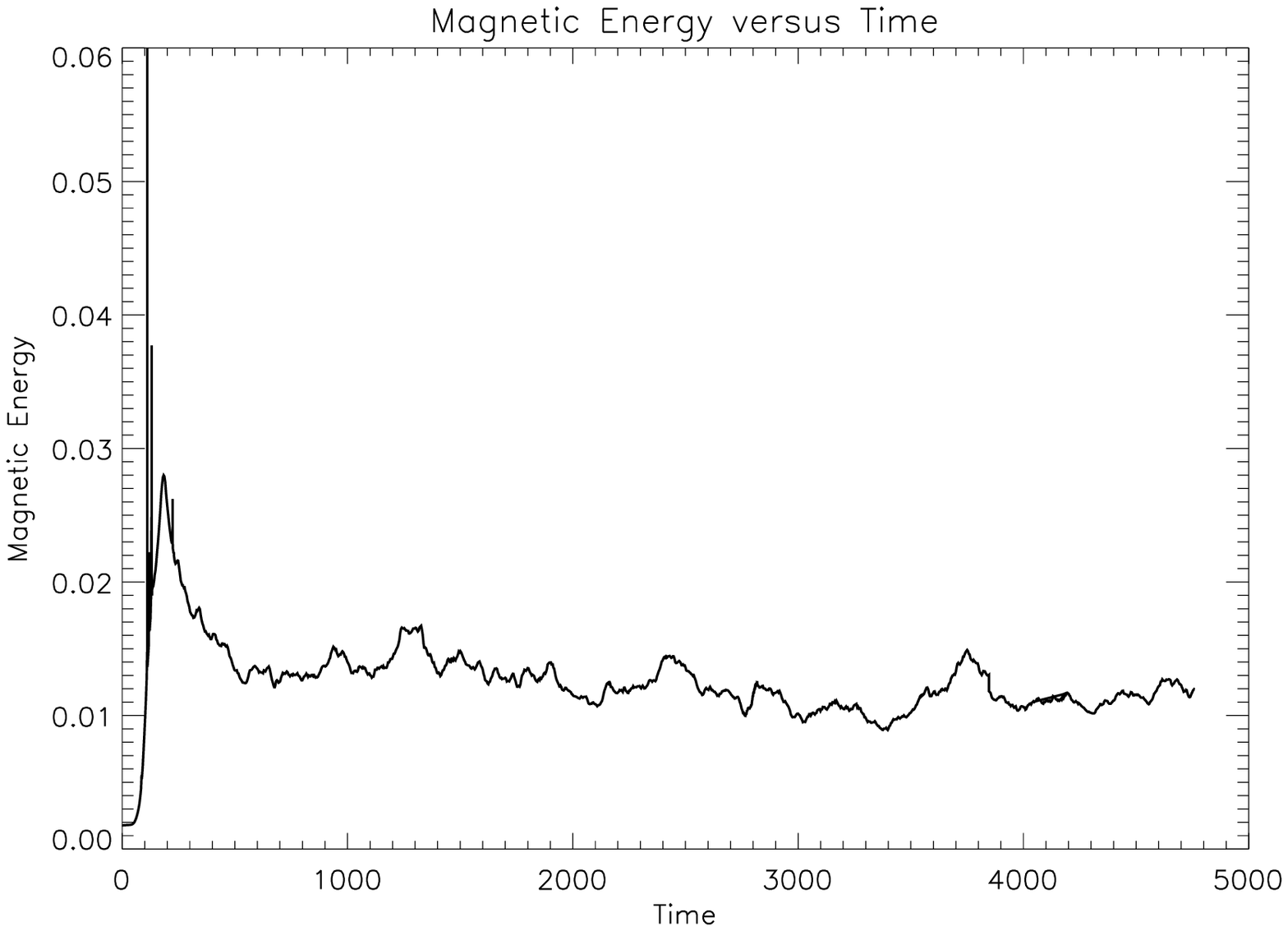,width=\columnwidth}}
\caption[]
{Magnetic energy in the Keplerian domain expressed in units of
the volume integrated pressure as a function of time for model E.}
\label{fig15}
\end{figure}

\begin{figure}
\centerline{
\epsfig{file=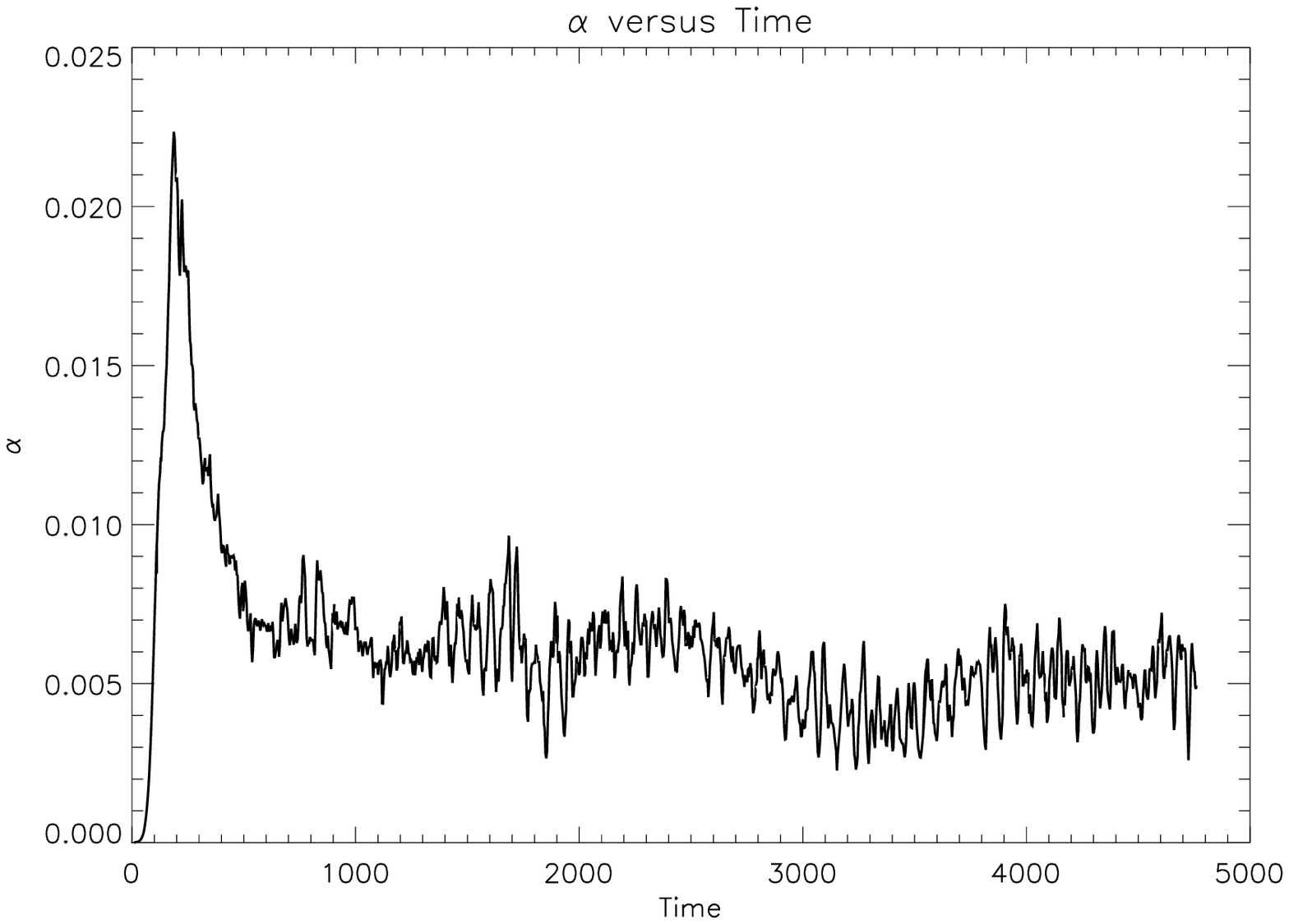,width=\columnwidth}}
\caption[]
{The stress parameter $\alpha$ volume averaged over the Keplerian domain
is plotted as a function of time for model E.}
\label{fig16}
\end{figure}

\begin{figure}
\centerline{
\epsfig{file=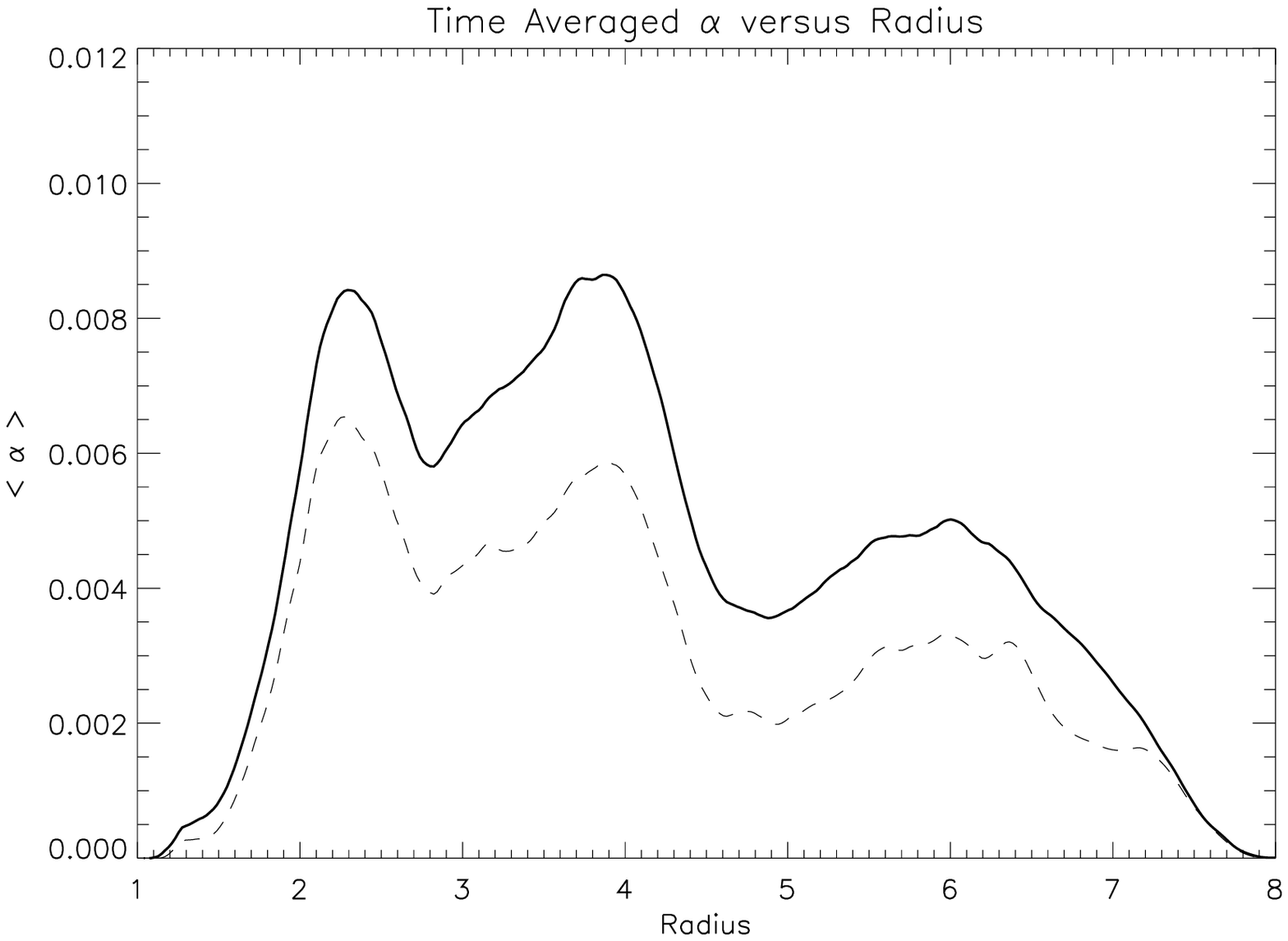,width=\columnwidth}}
\caption[]
{The time averaged stress parameter $\alpha$ as a  function of radius
for the model E.}
\label{fig17}
\end{figure}

This model was initiated with a poloidal magnetic field
and had a value of $c/(r \Omega)=0.1$ throughout the Keplerian domain.
The inner edge of the active Keplerian domain was located at
$r=1.2$ and the outer edge at $r=7.2$, with the azimuthal extent being
$\pi /3$, so this model has the
largest radial extent that we consider in this work. This model
is used in paper II to study the interaction between a MHD turbulent disc
and a giant protoplanet.

The temporal evolution of the magnetic energy in units of the volume
integrated pressure is shown in figure~\ref{fig15}.
This quantity saturates at a value of $\sim 0.01$ in the final
turbulent state, similar to the previous runs described. The volume averaged
stress parameter, $\alpha$, is plotted as a function of time in 
figure~\ref{fig16}, and shows similar behaviour to the previous runs described,
saturating at a value of $\sim 5 \times 10^{-3}$. The radial variation of the
time averaged stress parameter $\alpha$ is shown in figure~\ref{fig17},
with the solid line denoting the total $\alpha$, and the dashed line
denoting the magnetic contribution. The process of time averaging was initiated
at a time $t=3825.5$ and completed at $t=4244.1$. This time interval
corresponds to 66.7 orbits at $r=1$ and 3.5 orbits at $r=7.2$.

\begin{figure}
\centerline{
\epsfig{file=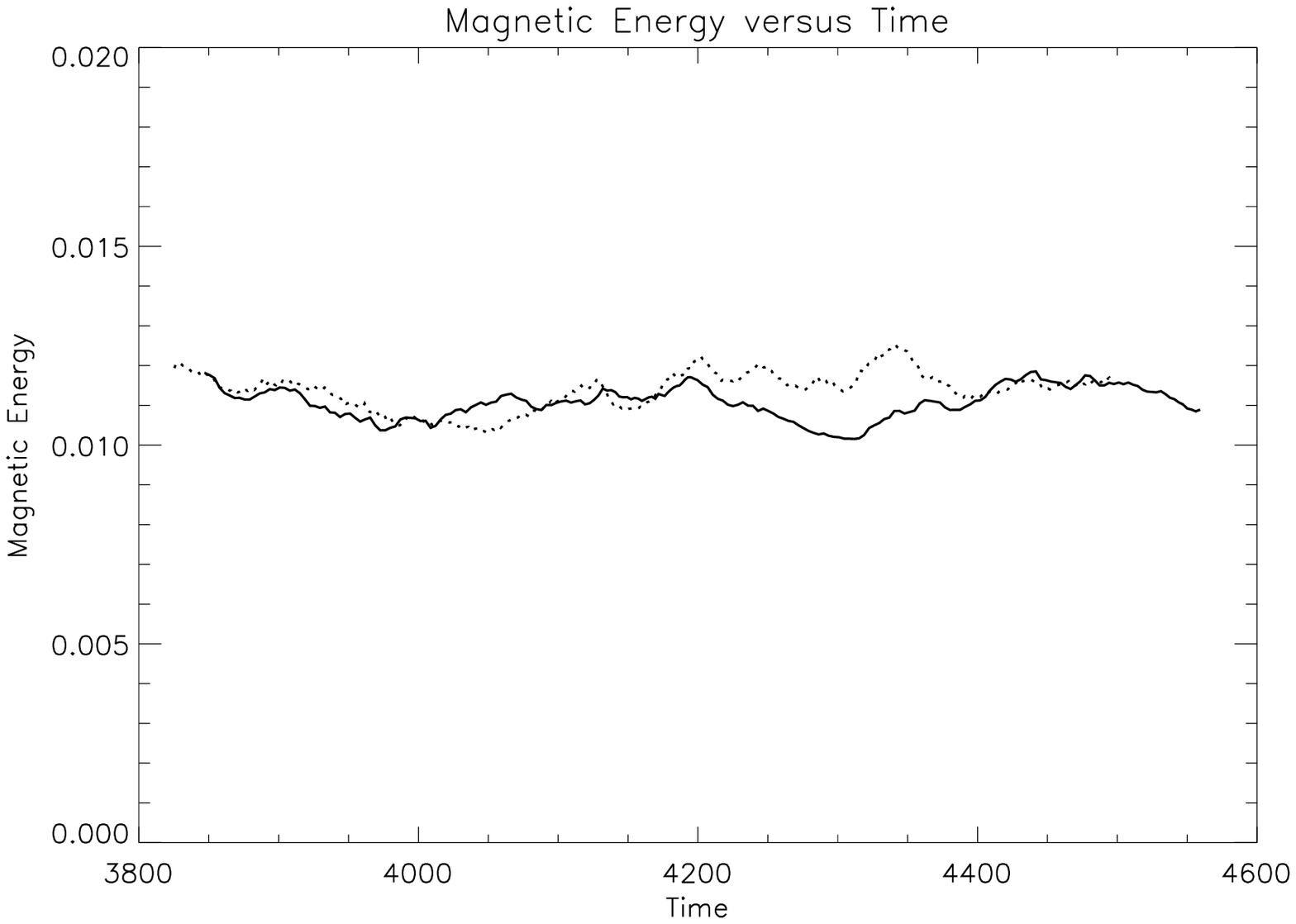,width=\columnwidth}}
\caption[]
{Magnetic energy in the Keplerian domain expressed in units of
the volume integrated pressure as a function of time for model E.
The solid line denotes the calculation performed in the inertial frame,
the dotted line denotes that calculated in the rotating frame.}
\label{fig18}
\end{figure}

\begin{figure}
\centerline{
\epsfig{file=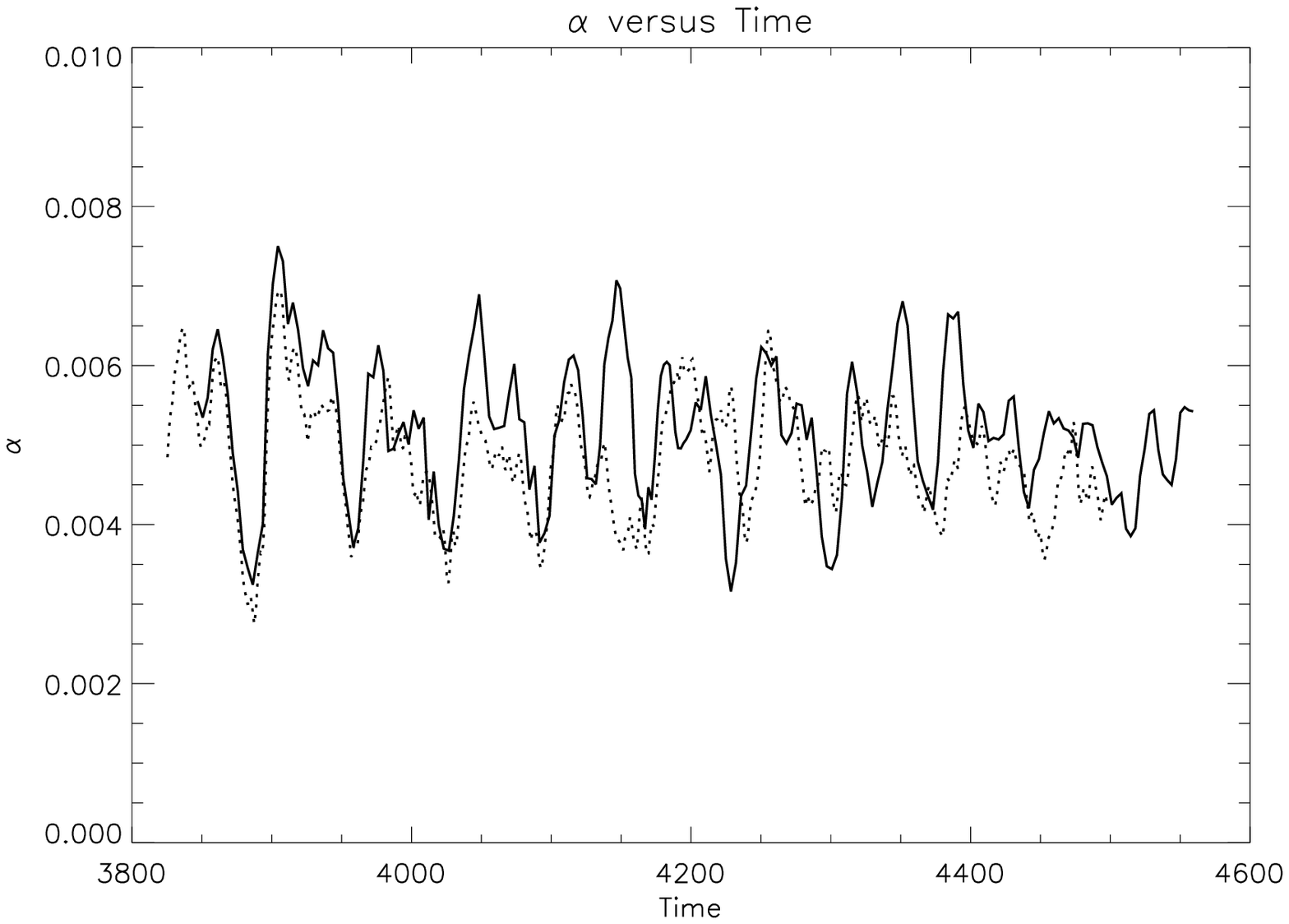,width=\columnwidth}}
\caption[]
{The volume averaged stress parameter $\alpha$ as a function of time
for the model E. The solid line denotes the calculation performed in the 
inertial frame, the dotted line denotes that calculated in the rotating frame.}
\label{fig19}
\end{figure}

At a time of $t=3825.5$, the disc model E was used to construct
a model with an azimuthal extent of $2 \pi$ for the purpose of studying
disc-planet interactions, (whilst the original model E run
was continued until $t=4500$).
This was achieved by simply patching six identical
copies of the $\pi/3$ model together to make a full $2 \pi$ disc, ensuring
of course that the azimuthal periodicity condition for the $\pi/3$ models were
properly utilised to construct the $2 \pi$ disc. In addition, the disc model
was transformed from being in an inertial frame to a rotating frame with
$\Omega_p=0.30645$. This value of $\Omega_p$ corresponds to the angular
velocity of material in circular Keplerian orbit at a radius of $r=2.2$.
We have checked that transforming the models from an inertial to a rotating
frame, and/or increasing the azimuthal extent of a model by patching
together identical copies, have no significant effect on the 
statistical properties of the turbulence. Figures~\ref{fig18} and \ref{fig19}
show the magnetic 
energy (in units of the volume integrated pressure) and the volume integrated
stress parameter $\alpha$ as functions of 
time for model E when transformed into the 
rotating frame (dotted line) and in the inertial frame (solid line).
Small divergences in the numerical values are observed, since the code is
now evolving a different {\em numerical} solution, but the statistical
properties remain very similar. An almost identical situation arises when
increasing the azimuthal domain from $\pi/3$ to $2 \pi$.

\begin{figure}
\centerline{
\epsfig{file=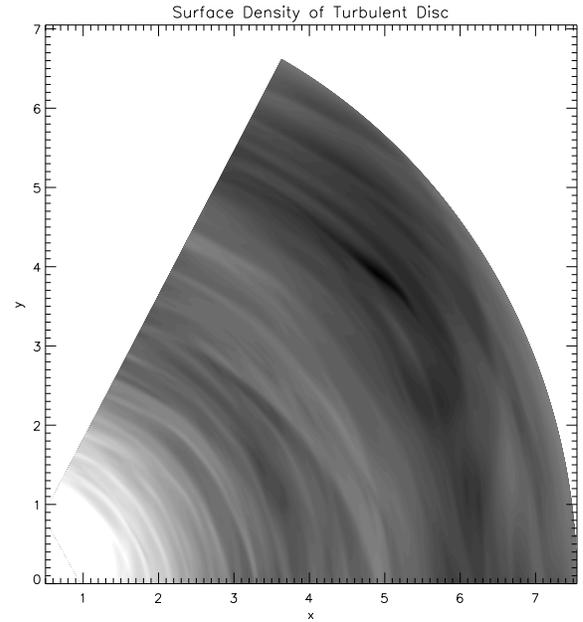,width=\columnwidth}}
\caption[]
{This plot shows the variations in density at the vertical midplane 
the disc model E.}
\label{fig20}
\end{figure}

\begin{figure}
\centerline{
\epsfig{file=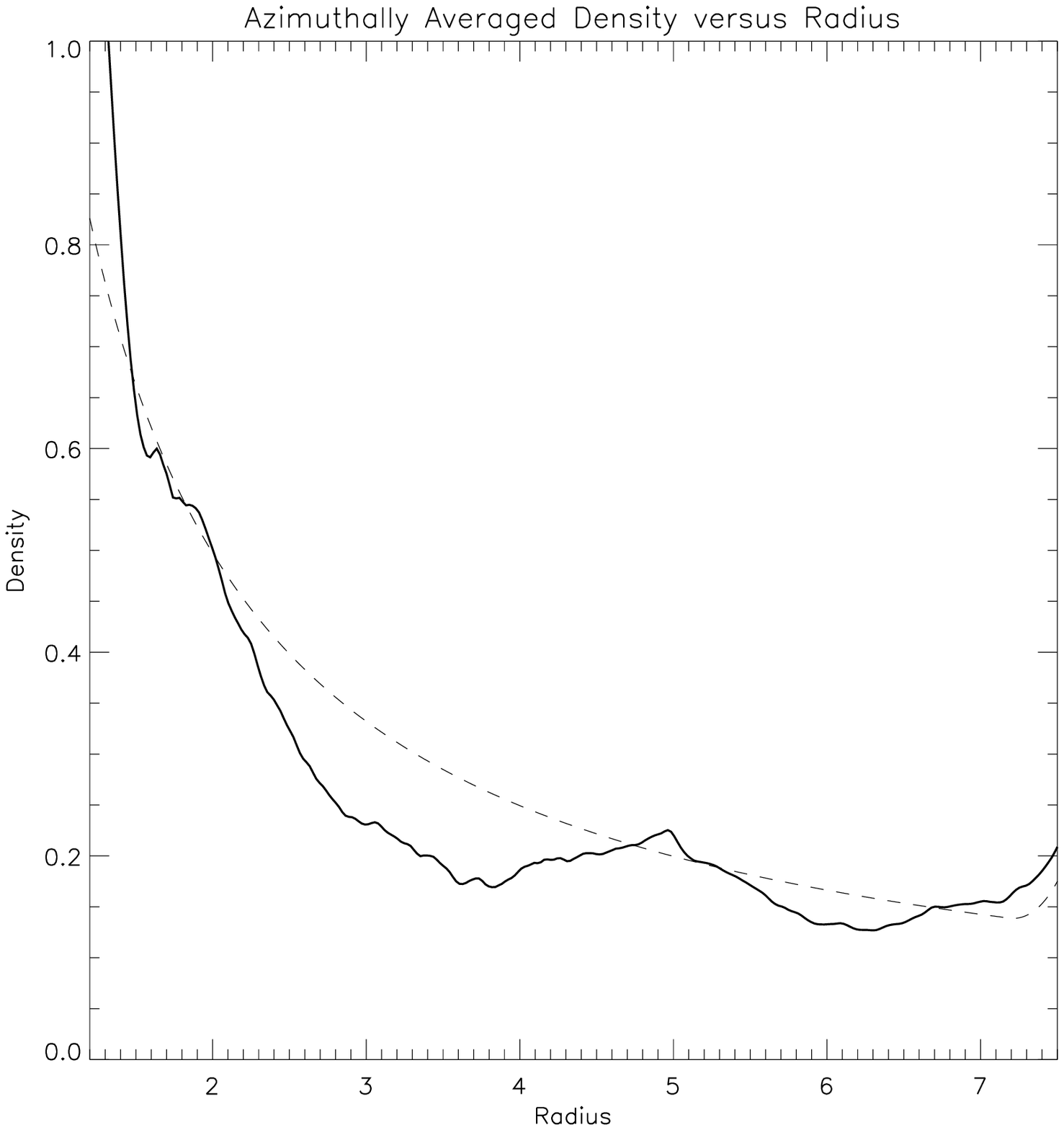,width=\columnwidth}}
\caption[]
{This plot shows the radial variation of the azimuthally averaged density 
at the 
vertical midplane of the disc model E.
The solid line corresponds
to time $t=3825.5$, whereas the dashed line shows
the initial values at $t=0$.}
\label{fig21}
\end{figure}

A greyscale plot of the density at the vertical midplane of the disc
is shown in figure~\ref{fig20} corresponding to a time $t=3825.5$. The usual
trailing waves generated by the turbulence can clearly be seen in this figure.
The variation of the azimuthally averaged midplane density as a function
of radius is plotted in figure~\ref{fig21}. The solid line corresponds
to the density distribution at time $t=3825.5$, whereas the dashed line shows
the initial values at $t=0$. It is clear from this figure that some
inward mass accretion has occurred.

\section{Discussion}\label{S6}

We have presented a study of cylindrical disc models
in which  a central domain in Keplerian rotation
is unstable to the MRI.  Models of varying disc size
and aspect ratio $H/r$ were considered.  They were initiated
with small scale
magnetic fields with zero net flux. Conservation of poloidal and toroidal
flux ensures that this situation is maintained throughout the simulations.
Input of flux through the boundaries which remain magnetically
inactive does not occur.
The models all attain a  turbulent state,    with statistical properties
that do not depend on the initial conditions which is expected
for genuine dynamo action. 

As these models have been prepared for studies of disc--
protoplanet interaction we have focused on relating the properties
of the turbulent models to classical viscous disc theory
[Shakura \& Sunyaev (1973)].  This is an important issue,
because  besides providing a conceptual framework
for disc studies, as emphasized by Balbus \& Papaloizou (1999)
it is still the main contact
between disc theory and observations. 

All models were found to attain a turbulent state with 
volume averaged stress parameter $\alpha \sim 5\times 10^{-3}$
and mean $\beta^{-1}\sim 0.01.$
We also found that the same results were obtained in a rotating frame
and in agreement with Hawley (2000) they were independent
of the extent of the azimuthal domain when this exceeded $\pi/3.$

The vertically and azimuthally averaged stress parameter
showed large radial fluctuations.
Time averaging for a period exceeding $3$ orbital
periods was found to  significantly reduce  them.
For models with $H/r =0.1$ stable variations with radius
 of a factor of two were then noted
whereas for models with $H/r =0.2$ 
 less variation was seen.
Variations  in the time averaged quantities
are most probably due to some memory of initial conditions
which take up to a viscous time to eradicate, this being shorter
for the thicker disc models.
The higher resolution thicker disc model tended to have
larger values of $\alpha$ than a lower resolution counterpart,
which may reflect a residual dependence
on resolution (see Brandenburg et al. 1996).

The vertically and azimuthally averaged radial velocity showed
large radial and   temporal fluctuations of up to two
orders of magnitude larger than  the inflow velocity expected from
classical viscous disc theory (see also SP).
 Time averaging for a period of at least $7-8$ orbital periods
at the outer boundary of the Keplerian domain was required to 
 reveal values of a magnitude comparable to the expected viscous inflow velocity.
Comparison with the value derived from the averaged stress 
using viscous disc theory yielded schematic agreement
for feasible averaging times.
It is likely that very long averaging times are needed to eliminate
residual fluctuations in the mean radial velocity  and that   such
an averaging operation may only be possible
for a very quiet and thin disc that has relaxed  to a   long term
statistical
steady state.

The behaviour described above must be borne in mind when
considering laminar disc simulations with anomalous Navier--Stokes viscosity
[e.g. Bryden et al (1999); Kley (1999); Lubow, Seibert \& Artymowicz (1999);
 D'Angelo  et al (2002)].
There, radial inflow velocities produced by the viscosity
produce phenomena like mass flow through gaps when a protoplanet is embedded
in a disc, due to the viscosity acting as a constantly acting
source of friction.
From the above discussion we might expect different
dynamical behaviour in the gap region
induced by a protoplanet in a genuinely turbulent disc (see paper II),
where the instantaneous 
velocity fluctuations are much larger than the time averaged
radial velocities arising from the turbulence--induced angular momentum 
transport.
More generally the classical viscous disc theory
cannot apply
to a disc undergoing rapid changes
due to external perturbation. It can only be used to describe the
parts of the disc that are in a quasi-steady condition for long enough
for appropriate averaging to be carried out.

\subsection{Acknowledgments} 

We acknowledge John Hawley, Steve Balbus, Adriane Steinacker, and
Caroline Terquem for useful
discussions. The computations reported here were performed using 
the UK Astrophysical
Fluids Facility (UKAFF) and the GRAND consortium supercomputing facility.


{}
\end{document}